\documentclass{PoS}

\title{Difficulties in Probing Nuclear Physics:  A Study of $^{44}$Ti and $^{56}$Ni}

\ShortTitle{Probing Nuclear Physics}

\author{
\speaker{Aimee Hungerford$^{ab}$},
Christopher L. Fryer$^{ab}$, 
Francis X. Timmes$^{ac}$,
Patrick Young$^{ac}$, 
Michael Bennett$^{ad}$,
Steven Diehl$^{abe}$,
Falk Herwig$^{adg}$,
Raphael Hirschi$^{ad}$,
Marco Pignatari$^{adg}$,
Georgios Magkotsios$^{acg}$,
and Gabriel Rockefeller$^{ab}$ \\
\llap{$^a$}The NuGrid Collaboration\\
\llap{$^b$}Computational Methods (CCS-2), Los Alamos National Laboratory, Los Alamos, NM, 87544, USA\\
\llap{$^c$}School of Earth and Space Exploration, Arizona State University, Tempe, AZ 85287, USA\\
\llap{$^d$}Astrophysics Group, Keele University, ST5 5BG, UK\\
\llap{$^e$}Theoretical Astrophysics Group (T-6), Los Alamos National Laboratory, Los Alamos, NM, 87544, USA\\
\llap{$^f$}Dept. of Physics \& Astronomy, Victoria, BC, V8W 3P6, Canada\\
\llap{$^g$} Joint Institute for Nuclear Astrophysics, University of Notre Dame, IN, 46556, USA\\
E-mail:\email{fryer@lanl.gov}
}

\abstract{The nucleosynthetic yield from a supernova explosion depends
upon a variety of effects: progenitor evolution, explosion process,
details of the nuclear network, and nuclear rates.  Especially in
studies of integrated stellar yields, simplifications reduce these
uncertainties.  But nature is much more complex, and to actually 
study nuclear rates, we will have to understand the full, complex 
set of processes involved in nucleosynthesis.  Here we discuss a 
few of these complexities and detail how the NuGrid collaboration 
will address them.}

\FullConference{10th Symposium on Nuclei in the Cosmos\\
		 July 27 - August 1 2008\\
		 Mackinac Island, Michigan, USA}

\begin{document}

\section{Understanding Key Rates in Astrophysics}

\begin{figure}
\center{\includegraphics[width=0.8\textwidth]{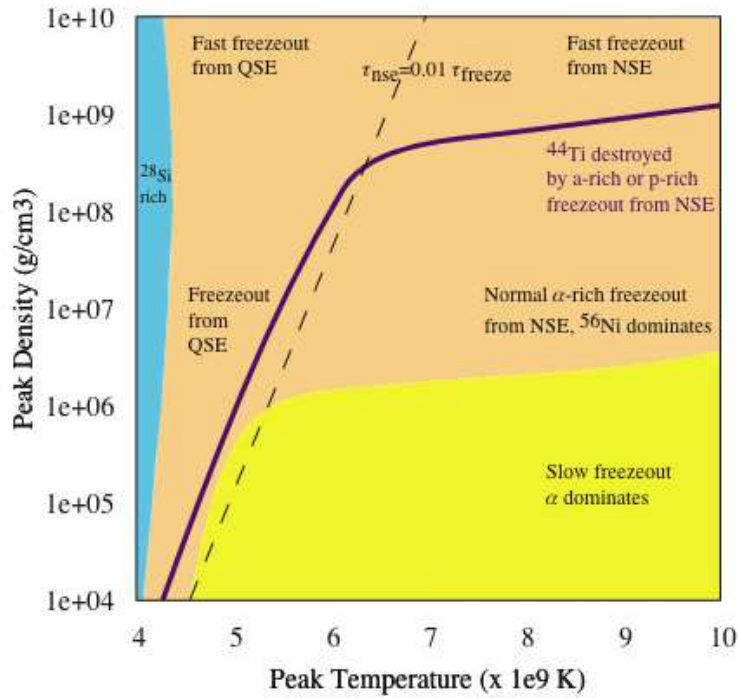}}
\caption{A cartoon displaying the key effects in nucleosynthesis for
shocked elements focusing on the $^{44}$Ti and $^{56}$Ni
yields~\cite{Hun08}.  Note that the position of the particles
determines which effects (fast freezeout, slow freezeout,
$\alpha-$rich freezeout) occur and this determines the important rates
for these yield calculations.  The position of the particle used in
the The et al.~\cite{the98} study ($\rho=10^7 {\rm g cm^{-3}}, T=5.5
\times 10^9 {\rm K}$) led them to believe one rate was important, but
in reality, the trajectories vary tremendously.}
\label{fig:cartoon}
\end{figure}

The field of nuclear astrophysics is predicated on the belief that
astronomers can cull from the tens of thousands of rates a handful of
critical rates that define the nuclear yields in astronomy.  Although
at some level, this is true: rates at some critical weighting points
do make a big difference in the yields; for the most part,
complications in nucleosynthesis make it very difficult to pick out a
single rate.  Early astrophysics success in pinpointing specific rates
has driven the nuclear physics community to expect that if they solve
the rates surrounding a few tens of isotopes, they can solve nuclear
astrophysics.  But many of these successes that pinpointed specific
rates did so because they focused on very specific points in the
density/temperature evolution.  In nature, the rate pinpointed by
these studies may be important for only a small amount of material
and, when comparing to observations, they may be completely neglible.

Here we present some of the pitfalls that can occur in determining key
rates using our study of the production of $^{44}$Ti and $^{56}$Ni as
an example.  But the complexity of understanding the role nuclear
rates plays on nucleosynthesis spans all discussions of nuclear
astrophysics and we present an r-process example as well.  Finally, we
conclude with the approach that will be taken by the NuGrid team.

\section{Understanding $^{44}$Ti and $^{56}$Ni Production}

One example of the complexitiies in understanding nucleosynthesis is
the study of $\alpha$-element production and, in particular, the
production of $^{44}$Ti and $^{56}$Ni.  Let's make the simplifying
assumption that the yield of a piece of matter is determined solely by
its peak temperature and density.  Figure 1 is a cartoon of the peak
density/temperature space generally studied in explosive
nucleosynthesis.  The et al.~\cite{the98} did exactly this analysis,
focusing on a single peak temperature/peak density: ($\rho=10^7 {\rm g
cm^{-3}}, T=5.5 \times 10^9 {\rm K}$).  This density/temperature pair
lies directly on the boundary of two different effects.  As such, a
single rate might change the yield by a large amount and The et
al.~\cite{the98} found that the triple-$\alpha$ rate changed the yield
dramatically.  But elsewhere on this diagram, the triple-$\alpha$ rate
is unimportant.  Unless we can assume all explosions produce elements
only at a single point, studying that single point will provide us with
a skewed set of important rates.

\begin{figure}
\center{\includegraphics[width=0.7\textwidth]{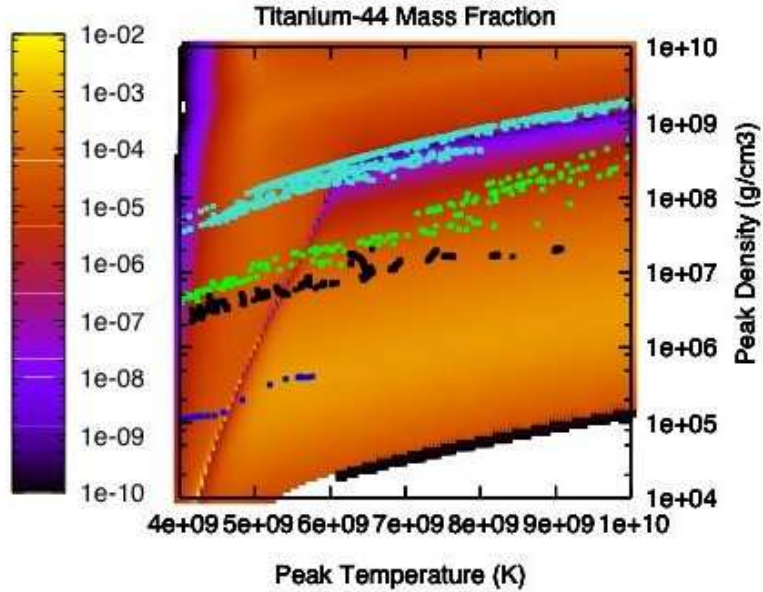}}
\caption{A plot of $^{44}$Ti yield on a peak density/temperature grid
with points from simulated explosions showing where they lie on this
grid.  From top to bottom, the points correspond to a
magnetohydrodynamic explosion of a collapsar
model~\cite{RFL08,rock08}, a rotating 2-dimensional
explosion~\cite{FH00}, and 2 weak-strong models mimicking fallback
gamma-ray bursts~\cite{FHY07}.  Note that the peak values span the
entire trajectory space.}
\label{fig:points}
\end{figure}

We have begun a more systematic study of this entire grid.  A first
step might be to determine what peaks are common in supernova
explosions.  Figure 2 again shows a plot of our peak
density/temperature grid with overlying points for 4 different
explosion calculations.  As one can see, they span a wide part of this
grid.  Not only are the points from different explosion models spread
in peak temperatures and densities, the points within each explosion
model possess a range of electron fractions for the fluid element
represented.  Magkotsios et al. 2008 provide a more detailed view of
the added complication from variation in electron fraction (these
proceedings.)  It appears that the supernova conditions will not
permit a narrowing of the important parameter space and an
understanding of the entire space is ultimately needed.

\begin{figure}
\center{\includegraphics[width=0.65\textwidth]{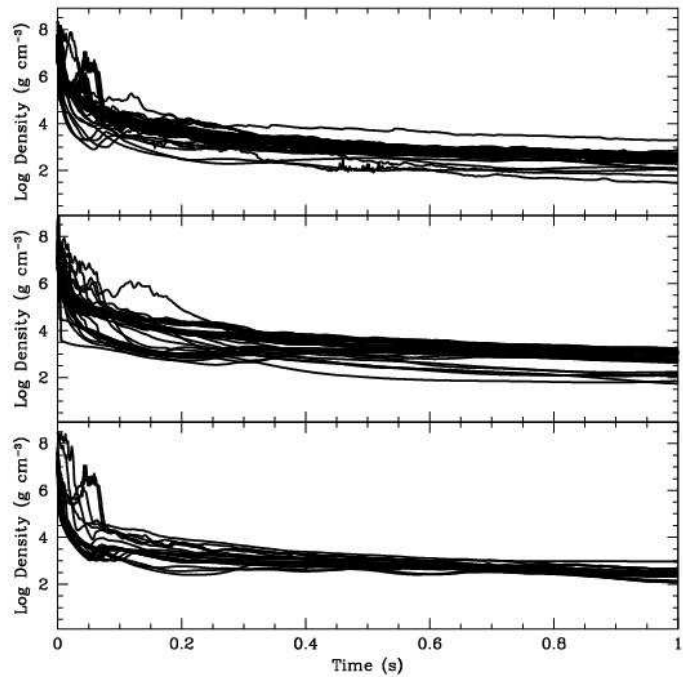}}
\caption{Density versus time for 3 sets of particles:  
high entropy (top), low entropy (middle) and particles that produced 
reasonable amounts of third r-process peak isotopes (bottom).  The 
zero point on the time axis is set to the time when the density 
reaches its maximum value (generally corresponding to the peak 
temperature as well).  It is very difficult to distinguish the 
peak densities from each other and it has not yet been determined 
what path is required to make the r-process.}
\label{fig:rho}
\end{figure}

With our simplifying assumption that we can determine everything from
a single peak density/temperature pair, the problem of determining a
yield (and the most crucial rates for that yield) presents us with
considerable work studying the entire grid space.  But, for some
problems, the work doesn't end there.  Our simplifying assumption is
not true for all nucleosynthetic problems.  In the study of nuclear
rates of r-process, most scientists have focused on the reactions and
trajectories behind wind-driven supernovae.  Again, this is a
too-narrow view and scientists working outside of this narrow view
have discovered an entirely new nucleosynthetic path (or paths) to
make r-process~\cite{mey02,FHHT1,FHHT2}.  Unfortunately, these new
paths depend on the subsequent evolution of the cooling matter as well
as the peak temperature/density.  Figure 3 shows density trajectories
for matter that did not make the r-proces peak and matter that
did~\cite{FHHT2}.  Matter with the same peaks produced very different
yields.  Even worse, it is not clear what trajectory is required to
make r-process isotopes.

\section{NuGrid Plans}

\begin{figure}
\includegraphics[width=0.5\textwidth]{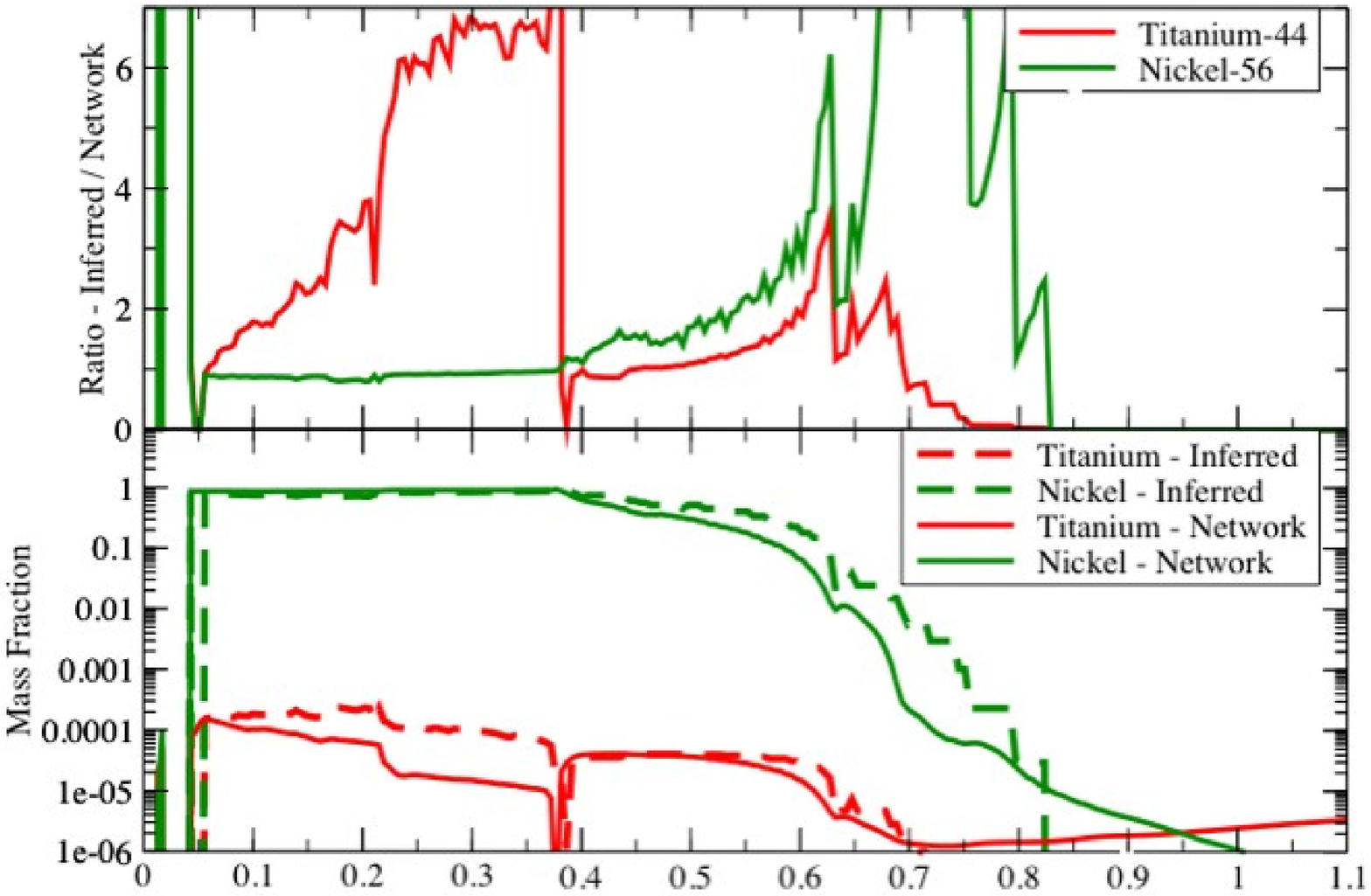}
\includegraphics[width=0.5\textwidth]{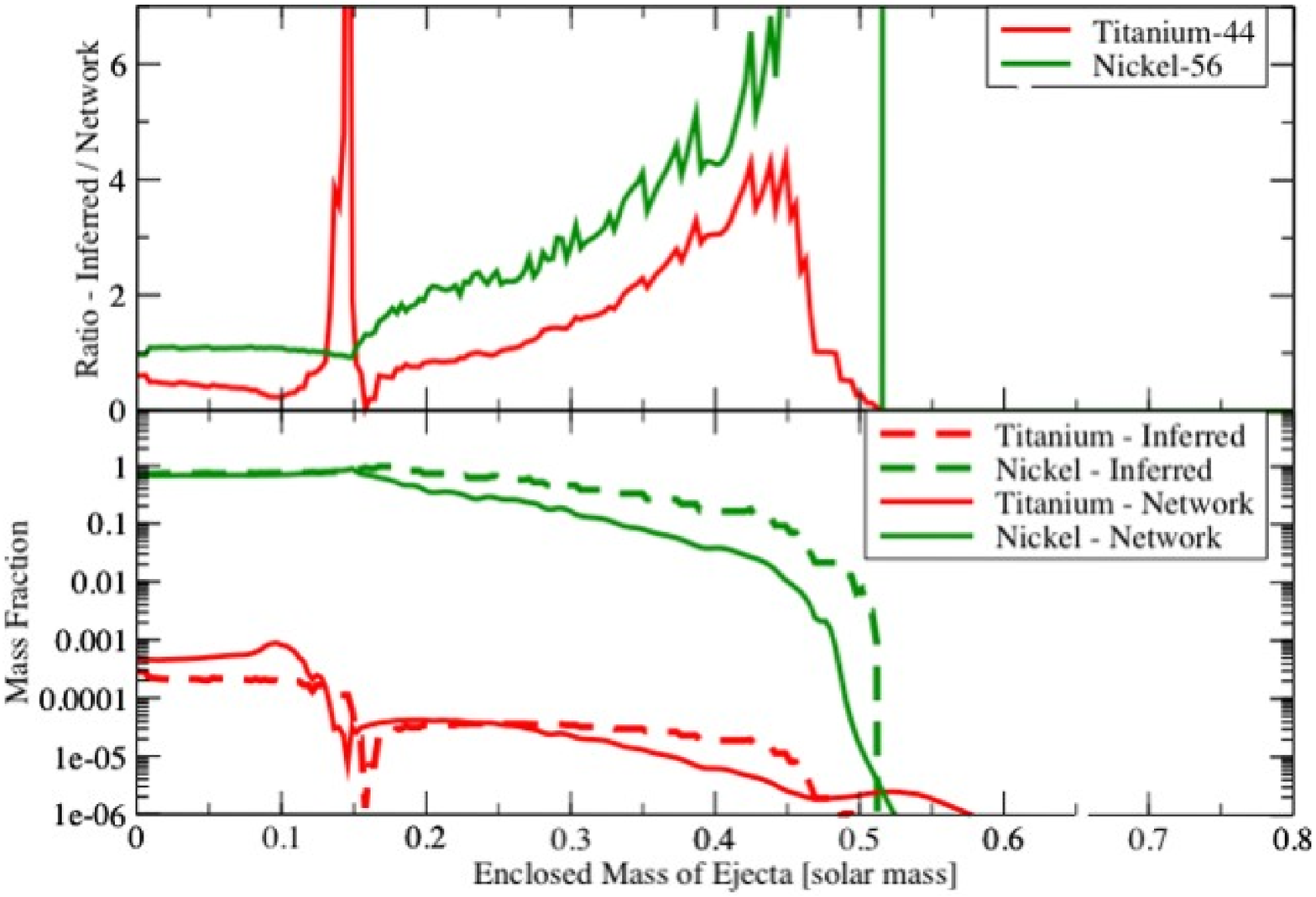}
\caption{$^{56}$Ni and $^{44}$Ti yields as a function 
of enclosed mass for two different stellar explosions.  
We compare the yields from the standard post-process 
network (solid lines) to those inferred using peak 
densities and temperatures.  The good agreement means 
that we can use these peak density/temperature 
diagrams to improve our intuition about nuclear 
network yields.}
\label{fig:traj}
\end{figure}

With all of these complexities, it would seem impossible to actually
understand astrophysical explosions and nuclear networks sufficiently
well to actually determine what rates are important.  But without
trying, we will definitely not solve this problem.  In many cases, the
peak temperature/density studies produce results that are very close
to studies that follow trajectories (Fig. 4) and we can use these
simple studies to develop our intuition.  But in the long run, we'll
have to approach this from all angles: studies of simplified problems,
like the density/temperature peak diagrams and their production
tracks, studies of temperature/density evolution tracks to better
understand which tracks produce what matter, and finally, integrated
yield studies (the more common study) to compare to observations.  One
approach alone will not work.  NuGrid is developing a suite of tools
ideally suited for all these studies and our collaboration will
approach this problem from all directions.

\end{document}